\documentclass[12pt]{article}
\usepackage[english,german,french,polish]{babel}
\usepackage[T1]{fontenc}
\usepackage{amsfonts}
\begin{document}

\baselineskip 0.75cm
\topmargin -0.6in
\oddsidemargin -0.1in

\let\ni=\noindent

\renewcommand{\thefootnote}{\fnsymbol{footnote}}

\newcommand{\CKM}{Cabibbo--Kobayashi--Maskawa }

\newcommand{\SM}{Standard Model }

\pagestyle {plain}

\setcounter{page}{1}

\pagestyle{empty}

~~~

\begin{flushright}
IFT--02/41
\end{flushright}

{\large\centerline{\bf ChroNodynamics or a quantum theory}}
{\large\centerline{\bf involving deviations from uniform run of time.}}

\vspace{0.3cm}

{\centerline {\sc Wojciech Kr\'{o}likowski}}

\vspace{0.2cm}

{\centerline {\it Institute of Theoretical Physics, Warsaw University}}

{\centerline {\it Ho\.{z}a 69,~~PL--00--681 Warszawa, ~Poland}}

\vspace{0.3cm}

{\centerline{\bf Abstract}}

\vspace{0.2cm}

Let us imagine that there is an overall quantum theory  (not necessarily recognized yet) of matter and energy ({\it i.e.}, of elementary fermions and bosons) interacting with the physical spacetime (treated on a quantum level). Since states of quantum spacetime are so far not observed directly, they ought to be projected out from the overall Hilbert space (much like states of a quantum medium in the optical model often constructed in nuclear physics). Then, in the reduced Hilbert space only states of quantum matter and energy are left, but now endowed with the energy width that enters through an antiHermitian interaction-like operator, a remainder of their coupling to the quantum spacetime. We postulate that such an energy width involves an averaged coupling of quantum matter and energy to a classical field of time deviations from the uniform time run (in the classical spacetime of special relativity). The well known time-temperature analogy helps us to fix other postulates leading altogether to a quantum theory we call chronodynamics (a loose analogue of thermodynamics of small deviations from thermal equilibrium).

\vspace{0.2cm}

\ni PACS numbers: 11.10.Lm , 11.90.+t , 12.90.+b .

\vspace{0.6cm}

\ni December 2002

\vfill\eject

~~~
\pagestyle {plain}

\setcounter{page}{1}

%\vspace{0.2cm}

\ni {\bf 1. Introduction}

\vspace{0.2cm}

It is a part of common knowledge in the quantum physics that there is a kind of correspondence between temperature and time. Speaking more precisely, in the quantum physics an analogy exists between the thermal-equilibrium operator $\exp(- H/kT)$ and the time-evolution operator $\exp(- iHt/\hbar)$, both generated by the Hamiltonian $H$ of a quantum system [1]. Such an analogy implies the correspondence

\vspace{-0.2cm}

%rownanie 1
\begin{equation}
1/kT \leftrightarrow it/\hbar
\end{equation}

\ni that parallels the notion of uniform temperature $T$ of a quantum system (persisting in the thermal equilibrium with a thermostat) with the notion of uniformly running time $t$ ascribed to a quantum system (most correctly on the ground of special relativity, if the effects of general relativity can be neglected).

If in a system some deviations $\delta T(\vec{x},t)$ appear from the uniform temperature $T$ of a thermal equilibrium, they are conducted throughout the system according to the heat conductivity equation. One may ask the exciting question, as to whether in Nature there may appear also some deviations $\delta t(\vec{x},t)$ from the uniformly running time $t$ [2] (hopefully, different and larger than the effects of general relativity), and then, what may be the mathematical equation governing their propagation throughout the system.

In the present paper, {\it postulating} the possibility of such time deviations (not being the effect of general relativity), we will describe them by the multiplicative scalar field $\varepsilon(\vec{x},t)$:

\vspace{-0.2cm}
 
%rownanie 2
\begin{equation}
t + \delta t(\vec{x},t) \equiv [1 + \varepsilon (\vec{x},t)]t \;,\; \delta t(\vec{x},t) \equiv \varepsilon (\vec{x},t) t\,.
\end{equation}

\ni Since time $t$ is always a classical real parameter, the scalar $\varepsilon (\vec{x},t)$ ought to be a classical real field. Of course, the values of $\varepsilon(\vec{x},t)$ should be tiny: $ -1 \ll \varepsilon(\vec{x},t) \ll 1$. Note that the scalar $\varepsilon(\vec{x},t)$ is connected with the scaling field, but acting only for time $t$ (not for the space coordinates $\vec{x})$. As we argue in Section 3, this action can be made explicitly Lorentz covariant in a way {\it \`{a} la} Tomonaga-Schwinger.

\vspace{0.2cm}

\ni {\bf 2. Classical field equation for time deviations}

\vspace{0.2cm}

We will assume the field equation for $\varepsilon(x) \equiv \varepsilon(\vec{x},t)$ in the form of a relativistic differential equation that in the formal nonrelativistic approximation becomes a {\it conductivity equation}. Such a form can be obtained, indeed, from the {\it tachyonic-type} Klein-Gordon equation which, in the sourceless case, may be written as

%rownanie 3
\begin{equation}
\left( \Box + \frac{1}{\lambda^2}\right) \varepsilon^{(0)} (x) = 0 \,,
\end{equation}

\ni where $\Box = -\partial^2 = \Delta - (1/c^2) \partial^2/\partial t^2$ with $\partial_\mu = \partial/\partial x^\mu$ and $x = (x^\mu) = (ct, \vec{x})$. In fact, inserting $\varepsilon^{(0)}(x) = \phi(x) \exp(ct/\lambda)$, where $\lambda = \sqrt{\lambda^2} >0$ is a lengthdimensional constant, we calculate from Eq. (3):

%rownanie 4
\begin{equation}
0 = \left( \Box + \frac{1}{\lambda^2}\right) \varepsilon^{(0)}(x) = e^{\frac{ct}{\lambda}}\left( \Box -  \frac{2}{\lambda c}\frac{\partial}{\partial t}\right) \phi(x) \simeq e^{\frac{ct}{\lambda}}\left( \Delta -  \frac{2}{\lambda c}\frac{\partial}{\partial t}\right) \phi(x)  \,.
\end{equation}

\ni Here, the last step is valid in the nonrelativistic approximation when, in addition, the constant $\lambda$ is such that

%rownanie 5
\begin{equation}
|\frac{1}{c^2} \frac{\partial^2}{\partial t^2} \phi(x)| \ll \frac{2}{\lambda c}|\frac{\partial}{\partial t}\phi(x)|\;.
\end{equation}

\ni Then, from Eq. (4) the conductivity equation for $\phi(x)$ with a lengthdimensional conductivity coefficient $\lambda/2$,

%rownanie 6
\begin{equation}
\left(\Delta - \frac{2}{\lambda c} \frac{\partial}{\partial t} \right) \phi(x) = 0 \;,
\end{equation}

\ni follows approximately.

Notice that in the case of conventional Klein-Gordon equation $\left( \Box - 1/\lambda^2 \right) \varepsilon^{(0)} (x) = 0$ we should insert $\varepsilon^{(0)} (x) = \phi(x) \exp (-ict/\lambda)$ in order to cancel the term $-1/\lambda^2$ in the Klein-Gordon operator. Then, in the nonrelativistic approximation and with the use of Eq. (5) we would obtain the Schr\"{o}dinger equation $[\Delta + i(2/\lambda c) \partial/\partial t] \phi(x) = 0$, where $1/\lambda c = m/\hbar $.

The homogeneous tachyonic-type Klein-Gordon equation (3) allows for plane-wave solutions $\exp(-i k\cdot x)$ of two interesting kinds: the ultraluminal plane-wave solutions with $k_0 = \sqrt{\vec{k}^2 - 1/\lambda^2}$, where $|\vec{k}| > 1/\lambda$, and the damped-in-time plane-wave solutions with $k_0 = -i|k_0| = -i\sqrt{1/\lambda^2 - \vec{k}^2}$, where $|\vec{k}| < 1/\lambda$. In the second case, $\exp(-i k\cdot x) = \exp(i\vec{k}\cdot \vec{x} - c|k_0|t)$. The less interesting solutions with $k_0 = +i|k_0| = + i\sqrt{1/\lambda^2 - \vec{k}^2}$, where $|\vec{k}| < 1/\lambda$, also exist, leading to $\exp(-i k\cdot x) = \exp(i\vec{k}\cdot \vec{x} + c|k_0|t)$ {\it i.e.}, the growing-in-time plane-wave solutions.

Now, the crucial problem arises, what is the source for the scalar field $\varepsilon(x)$. Such a source should be introduced into Eq. (3) on its rhs. Since the scalar field $\varepsilon(x)$ is classical, its source ought to be also a classical scalar field. Connecting the time deviations described by $\varepsilon(x)$ with the state of matter and energy ({\it i.e.}, of fermions and bosons), it is natural to consider in the context of $\varepsilon(x)$ the classical or, more precisely, averaged four-current $ j^\mu (x)$ of all fermions and bosons appearing in Nature. Such a structure can be presented generically as

%rownanie 7
\begin{equation}
j^\mu (x) \equiv {\rm Tr}[\rho(t) J^\mu(\vec{x})]
\end{equation}

\ni or, for pure quantum states, as

%rownanie 8
\begin{equation}
j^\mu (x) \equiv \,<\!\psi(t) | J^\mu(\vec{x}) | \psi(t)\!>_{\rm av}\,,
\end{equation}

\ni where the Schr\"{o}dinger picture for quantum states and operators is used (the subscript av in Eq. (8) denotes averaging over spins of all particles present in the considered state). Here, $\rho(t)$ and $\psi(t)$ denote, respectively, the density matrix and the state vector describing alternatively the considered state of the system, while $J^\mu (\vec{x})$ stands for the operator of four-current of all fermions and bosons [represented in the Schr\"{o}dinger picture; if the Heisenberg picture is used, the four-current operator depends both on $\vec{x}$ and $t$: $J^\mu_H(x)$].

Since the classical field $\varepsilon(x)$ is a scalar, it is natural to expect that the scalar $\partial_\mu j^\mu (x)$, formed with the use of classical four-current (7) or (8), is the source of $\varepsilon(x)$. Thus, we will {\it postulate} that the inhomogeneous tachyonic-type Klein-Gordon equation of the form

%rownanie 9
\begin{equation}
\left( \Box + \frac{1}{\lambda^2}\right) \varepsilon (x) = -g_\varepsilon \lambda ^2 \frac{1}{c} \, \partial_\mu j^\mu (x) \,,
\end{equation}

\ni where $g_\varepsilon > 0 $ denotes a dimensionless coupling constant, is the classical field equation for the scalar $\varepsilon(x)$ (describing, as in Eq. (2), the time deviations $\delta t(x) \equiv \varepsilon (x) t$ from the uniform time run $t$). Note that in the source term on the rhs of Eq. (9) there figures  the constant $\lambda^2$ (the same as on its lhs) because of dimensional reasons. Evidently, the new coupling constant $g_\varepsilon $ should be tiny in order to generate tiny time deviations. Hopefully, its value will be at one time determined experimentally.

The field equation (9) tells us that $\varepsilon(x)$ is generated at such spacetime points $x$, where
$\partial_\mu j^\mu (x) \neq 0$ {\it i.e.}, where the averaged total number of particles, $N(t) = \frac{1}{c} \int d^3\vec{x} j^0(x) $, changes locally. Since the total number of fermions (more precisely, fermions minus antifermions) does not change at all (even locally), the changing part of $N(t)$ is given by the averaged total number of bosons {\it i.e.}, $\partial_\mu j^\mu (x) = \partial_\mu j_{\rm boson}^\mu (x)$ (this conclusion is true, if $<\!\psi(t)|\psi(t)\!> =$ const; strictly speaking, the conventional conservation of norm of the state vector will turn out an approximation only, but an extremely good one). Obviously, for the whole space

%rownanie 10
\begin{equation}
\frac{d N(t)}{dt} \equiv \int d^3\vec{x}\; \partial_{\,0}\, j^0(x) = \int d^3\vec{x}\,\partial_\mu\, j^\mu(x)\;, 
\end{equation}

\ni if the spatial current $\vec{j}(x)$ vanishes properly in the spatial infinity.

We would like to emphasize that, though in our experience on the Earth the physical systems emitting particles {\it e.g.} photons (for which $\partial_\mu j^\mu (x) > 0$) are rather exceptional requiring some excitations to be realized, in the cosmic scale they are the most frequent natural systems such as, for instance,  the Sun and other stars. On the other hand, systems absorbing particles {\it e.g.} photons (for which $\partial_\mu j^\mu (x) < 0$) appear often, even on the Earth. Thus, the situation where $\partial_\mu j^\mu (x) \neq 0$ is normal in the Universe.

\vspace{0.2cm}

\ni {\bf 3. Quantum state equation dependent on time deviations}

\vspace{0.2cm}

Thus, we have the classical field equation for $\varepsilon(x)$, determining time deviations in terms of local changes of averaged total number of particles in the quantum state described by $\rho(t)$ or $\psi(t)$. Now, we can pass to the second crucial problem, how these time deviations affect in turn the quantum time evolution of the density matrix $\rho(t)$ or state vector $\psi(t)$. This problem, if solved, closes the logical loop in the structure of a possible dynamical theory of time deviations that may be called {\it chronodynamics} (in order to stress its loose, but characteristic, analogy with the thermodynamics of small deviations from thermal equilibrium).

Of course, in absence of time deviations, the conventional density matrix equation [3]

%rownanie 11
\begin{equation}
i \hbar \frac{d\rho^{(0)}(t)}{dt} = [H\,,\,\rho^{(0)}(t)]
\end{equation}

\ni or state equation [3]

%rownanie 12
\begin{equation}
i \hbar \frac{d\psi^{(0)}(t)}{dt} = H\,\psi^{(0)}(t)
\end{equation}

\ni works (in the Schr\"{o}dinger picture). These equations should be modified, however, when some time deviations appear.

To this end, let us imagine our system of particles as an interacting subsystem of a larger quantum system whose other interacting subsystem is the physical spacetime treated dynamically on a quantum level (we assume here that such a quantum theory of spacetime exists potentially; its explicit structure will not be needed for our purposes). Note that for our original system of particles the physical spacetime plays the role of some unavoidable classical surroundings, where the classical field $\varepsilon(x)$ describing time deviations may propagate. Then, let us project out from the Hilbert space of the larger quantum system all states of the interacting subsystem  corresponding to the physical spacetime, leaving only states of the interacting subsystem identified with our system of particles. This projecting-out procedure answers the situation, where the potential quantum spacetime is not observed directly. For pure quantum states, such a procedure [4], analogical to the derivation of the simple optical model in nuclear physics, leads us to the state equation involving a new antiHermitian operator $-i\Gamma$ added to the Hermitian energy operator $H$, where $\Gamma$ is a Hermitian operator that may be called {\it energy width}. Thus, the modified state equation is

%rownanie 13
\begin{equation}
i \hbar \frac{d\psi (t)}{dt} = (H - i\Gamma)\psi (t)
\end{equation}

\ni (in the Schr\"{o}dinger picture). From the methodological point of view we must treat this quantum state equation for particles ({\it i.e.}, for matter and energy) as a {\it postulate}. Generically, we get the modified density matrix equation

%rownanie 14
\begin{equation}
i \hbar \frac{d\rho(t)}{dt} = [H,\rho(t)] - i\{\Gamma,\rho(t)\}
\end{equation}

\ni (in the Schr\"{o}dinger picture).

Above, we assume hopefully that the potentially existing quantum theory of spacetime stresses the role of time deviations among all, {\it a priori} possible, spacetime deviations, and that it works in the framework of special relativity (for the problem of its explicit covariance {\it cf.} a paragraph a bit later). This theory {\it in spe} is different from the long\-awaited quantum gravity, conventionally identified with general relativity in a quantum version. Note that deformations of space could be included into our chronodynamics, when $ x^\mu + \delta x^\mu(x) \equiv [1+\varepsilon(x)] x^\mu\;,\;\delta x^\mu(x) \equiv \varepsilon(x) x^\mu $ with $\varepsilon(x)$ related to the scaling field. In this paper we do not consider such an extension, believing in the existence of differences between deformations of time and space (considered on the ground of special relativity).

To complete the structure of chronodynamics we must construct the operator of energy width $\Gamma $, adequate to our situation. It seems natural to expect that this operator ought to involve the coupling of field $\varepsilon(x)$ to the source $\partial_\mu j^\mu (x)$ appearing in the field equation (9). In the present paper, we will {\it postulate} that 

%rownanie 15
\begin{equation}
\Gamma(t) \equiv {\bf 1} g_\varepsilon \hbar \int d^3\vec{x}\,\partial_\mu j^\mu(x) \varepsilon(x)\;, 
\end{equation}

\ni where $g_\varepsilon $ is the same tiny dimensionless coupling constant as that in the source term of field equation (9). The constant $\hbar$ appears because of dimensional reasons. The unit operator {\bf 1} in the Hilbert space of our system of particles guarantees in Eq. (13) the (trivial) operator character of $\Gamma (t)$. The rest of $\Gamma (t)$ is classical. Hereafter, by $\Gamma (t)$  we will understand this rest,

%rownanie 16
\begin{equation}
\Gamma (t) \equiv g_\varepsilon \hbar \int d^3\vec{x}\,\partial_\mu j^\mu(x) \varepsilon(x)\;.
\end{equation}

\ni Thus, the operator (15) appearing in Eq. (13) will be equal to ${\bf 1}\Gamma (t)$.

In the case of pure quantum states, the set of Eqs. (9) with (8) and (13) with (15) for the classical field $\varepsilon (x)$ and quantum state vector $\psi (t)$ defines a theory, consistent with special relativity, that we have called chronodynamics. This is an extension of the conventional quantum dynamics, when the time deviations $\delta t(x) \equiv \varepsilon (x) t$ are allowed in the reference system, where $\psi (t)$ is the state vector (and when the effects of general relativity can be neglected).

Note that both $\delta t(x)$ and $\psi (t)$ can be replaced by the explicitly Lorentz-covariant structures by means of the known Tomonaga-Schwinger construction [5, 6] using in the Minkowski space the spacelike three-dimensional hypersurface $\sigma (x)$ that replaces the running hyperplane $t =$ const. Then, the time deviations $\delta t(x) \equiv \varepsilon (x) t$ from $t$ are replaced by $\delta \sigma(x) \equiv \varepsilon (x) \sigma(x)$, the timelike deviations from $\sigma(x)$, and the function $\psi(t)$ by the functional $\psi[\sigma]$. As in the conventional theory without time deviations, the derivative $i \hbar d/dt $ is replaced by the functional derivative $i \hbar c\, \delta/\delta^4 \sigma(x)$, and the operator $H$, in our case $H - i \Gamma$, by its density. Here, $\delta^4 \sigma(x)$ in $\delta/\delta^4 \sigma(x)$ denotes the infinitesimal four-dimensional volume in the Minkowski space determined by the infinitesimal timelike deformation of the spacelike hypersurface $\sigma(x)$, localized around its point $x$. Since in a relativistic field theory the density of the operator $H_{\rm int} - i \Gamma $ with $H_{\rm int}$ denoting the interaction part of H is a Lorentz scalar, the Lorentz covariance of the Tomonaga-Schwinger form of state equation (13) becomes explicit in the interaction picture.

From Eq. (13) we obtain

%rownanie 17
\begin{equation}
\psi(t) = \psi^{(0)}(t) \exp\left[ -\frac{1}{\hbar} \int^t_{t_0} dt'\, \Gamma(t')\right] \;,
\end{equation}

\ni where $ \psi^{(0)}(t)$ satisfies the conventional state equation (12) with the initial condition  $ \psi^{(0)}(t_0) =  \psi(t_0)$ and $\Gamma(t)$ is given as in the definition (16). On the fundamental level, the Hamiltonian $H$ is always time-independent (in the Schr\"{o}dinger and  Heisenberg pictures). Then, Eq. (12) gives

%rownanie 18 
\begin{equation}
\psi^{(0)}(t) = \exp\left[ -\frac{1}{\hbar} H(t - t_0)\right] \psi^{(0)}(t_0)\;.
\end{equation}

\ni We can see from Eq. (17) that the norm of $\psi(t)$ changes in time, namely

%rownanie 19
\begin{equation}
<\psi(t)|\psi(t)> = \exp\left[ -\frac{2}{\hbar} \int^t_{t_0} dt'\, \Gamma(t')\right]
\end{equation}

\ni for $<\psi(t_0)|\psi(t_0)> = 1$. In contrast, $<\psi^{(0)}(t)|\psi^{(0)}(t)> = 1$. Naturally, the variation in the formula (19) is very slow, as $g_\varepsilon $ in Eq. (16) defining $\Gamma(t)$ is tiny. Generically, Eq. (14) gives

%rownanie 20
\begin{equation}
\rho(t) = \rho^{(0)}(t) \exp\left[ -\frac{2}{\hbar} \int^t_{t_0} dt'\, \Gamma(t')\right] \;,
\end{equation}

\ni where $\rho^{(0)}(t)$ fulfils the conventional density matrix equation (11) with the initial condition $\rho^{(0)}(t_0) = \rho(t_0)$. Hence,

%rownanie 21
\begin{equation}
{\rm Tr}\,\rho(t) = \exp\left[ -\frac{2}{\hbar} \int^t_{t_0} dt'\, \Gamma(t')\right] 
\end{equation}

\ni for Tr $\rho(t_0) = 1$. In contrast, Tr $\rho^{(0)}(t) = 1$. Note that in our case Eq. (14) takes the form
$i\hbar\, d\rho(t)/dt = [H,\rho(t)] - 2i\rho(t) \Gamma $ with $\Gamma(t)$ as defined in Eq. (16).

The formal reason for the variation in time of $<\!\psi(t)|\psi(t)\!>$ and, generically, of Tr~$\rho(t)$ is that the physical spacetime of chronodynamics is not included as a part into the considered quantum system in spite of mutual interactions exciting time deviations on the classical level. In fact, one can imagine that it is projected out (in the sense of its states) from the larger quantum system including it as an interacting subsystem on a quantum level [4].

In conclusion of the first part of this paper, we can see that Eqs. (9) [together with (8)] and (13) [together with (15)] form a mixed, classical-quantum set of two coupled equations for the classical field $\varepsilon(x)$ describing the time deviations $\delta t(x) \equiv \varepsilon(x) t$, and the quantum state vector $\psi(t)$ of time-evolving matter and energy. One may also say that this set is a quantum set of two coupled equations, namely the equation for the quantum state vector $\psi(t)$ in the classical (or, more precisely, classical-valued) field $\varepsilon(x)$ of time deviations introduced by means of $\Gamma (t)$, and the constraint imposed on the classical field $\varepsilon(x)$ dependent functionally on the quantum state vector $\psi(t)$ through $\partial_\mu j^\mu (x)$. Because of the bilinear form of $\psi(t)$ appearing in Eq. (8) this set of equations is (strictly speaking) nonlinear with respect to the state vector $\psi(t)$, what violates (slightly) the superposition principle for $\psi(t)$. This perturbs the fundamental probability interpretation of $\psi(t)$.

However, in the first-order perturbative approximation with respect to the tiny $g_\varepsilon $, where $\psi(t)$ in Eq. (8) is approximated by $\psi^{(0)}(t)$ satisfying the conventional state equation (12) (and $\varepsilon^{(0)}(x)$ fulfilling the homogeneous field equation (3) is put zero), the set of Eqs. (9) and (13) becomes linear with respect to $\psi^{(1)}(t)$, the state vector in the first-order perturbative approximation. In fact, in this approximation

%rownanie 22
\begin{equation}
\left( \Box + \frac{1}{\lambda^2}\right) \varepsilon^{(1)} (x) = -g_\varepsilon \lambda ^2 \frac{1}{c} \,\partial_\mu\, j^{\mu\,(0)}(x) 
\end{equation}

\ni with

%rownanie 23
\begin{equation}
j^{\mu\,(0)}(x) \equiv <\!\psi^{(0)}(t)|J^\mu (\vec{x})|\psi^{(0)}(t)\!>_{\rm av}\;,
\end{equation}

\ni and

%rownanie 24
\begin{equation}
i\hbar \frac{d\psi^{(1)}(t)}{dt} = \left[H - i{\bf 1} \Gamma^{(1)}(t) \right] \psi^{(1)}(t)
\end{equation}

\ni with

%rownanie 25
\begin{equation}
\Gamma^{(1)}(t) \equiv g_\varepsilon \hbar\int d^3\vec{x}\, \partial_\mu\, j^{\mu\,(0)}(x) \,\varepsilon^{(1)}(x)\;.
\end{equation}

\ni Here, $\varepsilon^{(0)}(x)$ fulfilling the homogeneous field equation (3) is put zero; then, the possible contribution $\partial_\mu j^{\mu\,(1)}(x) \varepsilon^{(0)}(x)$ with $j^{\mu\,(1)}(x) \equiv \,<\!\psi^{(0)}(t)| J^\mu(\vec{x})| \psi^{(1)}(t)\!>_{\rm av} +$ c.c. to the integrand in Eq. (25) is zero. Thus, in this approximation we obtain a classical-quantum set of two uncoupled equations for $\varepsilon^{(1)} (x)$ and $\psi^{(1)}(t)$. They give

%rownanie 26
\begin{equation}
\varepsilon^{(1)}(x) = -g_\varepsilon \lambda^2\,\frac{1}{c} \left( \Box + \frac{1}{\lambda^2}\right)^{-1} \partial_\mu j^{\mu\,(0)}(x)
\end{equation}

\vspace{-0.1cm}

\ni and

\vspace{-0.2cm}

%rownanie 27
\begin{equation}
\psi^{(1)}(t) = \psi^{(0)}(t) \exp\left[ -\frac{1}{\hbar} \int^t_{t_0} dt'\, \Gamma^{(1)}(t')\right] \;,
\end{equation}

\ni where $\psi^{(0)}(t_0) = \psi^{(1)}(t_0)$ and 

\vspace{-0.2cm}

%rownanie 28 
\begin{eqnarray}
\Gamma^{(1)}(t) & = & -g^2_\varepsilon \lambda^2 \frac{1}{c} \hbar \int d^3\vec{x}\,\partial_\mu j^{\mu\,(0)}(x) \left( \Box + \frac{1}{\lambda^2}\right)^{-1} \partial_\nu j^{\nu\,(0)}(x) \nonumber \\
& \equiv & -g^2_\varepsilon \lambda^2 \frac{1}{c} \hbar \int d^3\vec{x}\,q^{(0)}(x) \left( \Box + \frac{1}{\lambda^2}\right)^{-1} q^{(0)}(x)
\end{eqnarray} 

\vspace{-0.1cm}

\ni with

\vspace{-0.2cm}

%rownanie 29
\begin{equation}
q(x) \equiv \partial_\mu j^{\mu}(x)\;.
\end{equation}

\ni Due to the linearity of the state equation (24) with respect to $\psi^{(1)}(t)$ the state vector can be normalized at an initial moment and so, interpreted probabilistically, though its norm varies (very slowly) in time, namely

%rownanie 30
\begin{equation}
<\!\psi^{(1)}(t) | \psi^{(1)}(t)\!> = \exp \left[-\frac{2}{\hbar} \int^t_{t_0} dt'\, \Gamma^{(1)}(t') \right]
\end{equation}

\ni for $<\!\psi^{(1)}(t_0)|\psi^{(1)}(t_0)\!> = 1$. Note that in Eq. (25)

\vspace{-0.2cm}

%rownanie 31
\begin{eqnarray}
\partial_\mu j^{\mu\,(0)}(x)  & = & <\!\psi^{(0)}(t)| \frac{i}{\hbar\,c}[H, J^{(0)}(\vec{x})] + \vec{\partial}\cdot \vec{J}(\vec{x})| \psi^{(0)}(t)\!>_{\rm av}  \nonumber \\ & = & <\!\psi^{(0)}_H|\partial_\mu J^\mu_H(x)| \psi^{(0)}_H\!>_{\rm av}  \,,
\end{eqnarray} 

\ni where the label $H$ indicates the Heisenberg picture.

One should remember that, in chronodynamics, the deviation from the strict conser\-vation of norm for $\psi(t)$ has an optical-model-like character which is not of a fundamental nature, if only the "projected-out"~quantum level of the physical spacetime really exists and, in principle, can be observed (in a sense known from our quantum-dynamical experience). Note also that, correctly speaking, the optical-model-like connection of chronodynamics with the potential background theory including the quantum spacetime should be considered in the linear first-order perturbative approximation, where $\psi(t)$ of chronodynamics is replaced by $\psi^{(1)}(t)$ in the state equation (and by $\psi^{(0)}(t)$ in the averaged current of particles), since this background overall quantum theory is linear in the overall state vector. At the end of Section 6 an extreme case is considered, where the consistent background theory does not exist, but an act of abstraction referring to it is realized.

In the second part of this paper we discuss some simple consequences of chronodynamics.

\vspace{0.25cm}

\ni {\bf 4. Violation of optical theorem}

\vspace{0.25cm}

The first-order perturbative $S$ matrix in chronodynamics is related to the conventional $S$ matrix through the formula

%rownanie 32
\begin{equation}
S^{(1)} = S^{(0)}(t) \exp \left[-\frac{1}{\hbar} \int^{+\infty}_{-\infty} dt\, \Gamma^{(1)}(t) \right]\,,
\end{equation}

\ni where $\Gamma^{(1)}(t)$ given in Eq. (25) determines the first-order unitary defect, since $S^{(0)} S^{(0)\,\dagger} = {\bf 1} = S^{(0)\,\dagger} S^{(0)}$. In Eq. (32), the interval $-\infty,+\infty$ symbolizes a time interval  very long in comparison with a short reaction time (where $\Gamma^{(1)}(t) \neq 0$).

Thus, in place of the conventional optical theorem

%rownanie 33
\begin{equation}
\sigma^{(0)}_{{\rm tot}\,i} = \frac{(2\pi)^4\,\hbar^2}{v_i} \frac{1}{\pi} {\rm Im} R^{(0)}_{ii}
\end{equation}

\ni we obtain the modified formula [2]

%rownanie 34
\begin{equation}
\sigma^{(1)}_{{\rm tot}\,i} = \frac{(2\pi)^4\,\hbar^2}{v_i} \frac{1}{\pi} {\rm Im} R^{(1)}_{ii} \exp \left[-\frac{1}{\hbar} \int^{+\infty}_{-\infty} dt\, \Gamma^{(1)}(t) \right]\,.
\end{equation}

\ni Here, diagonal elements of the reaction matrices $R^{(0)}$ and $R^{(1)}$ appear, calculated in initial asymptotic states $i$ of two colliding particles (averaged over particle spins). Since

%rownanie 35
\begin{equation}
R^{(1)}_{ii} = R^{(0)}_{ii} \exp \left[-\frac{1}{\hbar} \int^{+\infty}_{-\infty} dt\, \Gamma^{(1)}(t) \right]\,,
\end{equation}

\ni the rhs of Eq. (34) includes the exponential squared.

\vspace{0.25cm}

\ni {\bf 5. Excitations of time deviations by a stationary source of particles}

\vspace{0.25cm}

Consider a pointlike stationary system, located at the point $\vec{x}_S$, emitting $1/\tau_S$ particles of all sorts per unit of time which form a constant averaged current $\vec{j}^{(0)}(\vec{x})$. It may be {\it e.g.} a source of light. According to Eq. (22) that takes here the form

%rownanie 36
\begin{equation}
\left(\Delta + \frac{1}{\lambda^2} \right) \varepsilon^{(1)}(\vec{x}) = -g_\varepsilon \lambda^2 \frac{1}{c} \,{\rm div}\, \vec{j}^{(0)}(\vec{x}) \;,
\end{equation}

\ni this system is also a pointlike source of the static field $\varepsilon^{(1)}(\vec{x})$ describing the time deviations $\delta t^{(1)}(\vec{x}) \equiv \varepsilon^{(1)}(\vec{x}) t$ thus excited in such a way. Assume for the source $ \vec{q}^{(0)}(\vec{x}) \equiv \,{\rm div}\, \vec{j}^{(0)}(\vec{x})$ of $ \varepsilon^{(1)}(\vec{x})$ a mathematical model that may be reasonably realistic and, at the same time, allows to solve Eq. (36) analytically:

%rownanie 37
\begin{equation}
{\rm div}\, \vec{j}^{(0)}(\vec{x}) \equiv \frac{1}{\tau_S}\left[ \delta^3(\vec{x}-\vec{x}_S) + \frac{\cos (|\vec{x}-\vec{x}_S |/\lambda)}{|\vec{x}-\vec{x}_S|^3} \sum_{l\, m_l} c_{l\, m_l} \frac{l(l+1)}{4\pi} Y_{l\,m_l}(\theta,\phi)\right]\,,
\end{equation}

\ni where $c_{l\, m_l} = c^*_{l\, -m_l}$ are dimensionless coefficients normalized in such a way that $\sum_{l\, m_l} c_{l\, m_l} Y_{l\, m_l}(0,0) =1$, while $\theta$ and $\phi$ denote the spherical angles of the vector $\vec{x}-\vec{x}_S$. Then, it follows from Eq. (36) that

%rownanie 38
\begin{equation}
\varepsilon^{(1)}(\vec{x}) = \frac{g_\varepsilon \lambda^2}{4\pi c \tau_S}\, \frac{\cos (|\vec{x}-\vec{x}_S| /\lambda)}{|\vec{x}-\vec{x}_S|}\sum_{l\, m_l} c_{l\, m_l} Y_{l\,m_l}(\theta,\phi)\,.
\end{equation}

\ni In particular, if $c_{l\, m_l} = 0$ for $l>0$ {\it i.e.}, ${\rm div}\, \vec{j}^{(0)}(\vec{x}) \equiv (1/\tau_S) \, \delta^3(\vec{x}-\vec{x}_S)$, then

%rownanie 39
\begin{equation}
\varepsilon^{(1)}(\vec{x}) = \frac{g_\varepsilon \lambda^2}{4\pi c \tau_S}\, \frac{\cos (|\vec{x}-\vec{x}_S| /\lambda)}{|\vec{x}-\vec{x}_S|}\,.
\end{equation}

In stationary situations, as {\it e.g.} for the system discussed here, the energy width (25) is time-independent,

%rownanie 40
\begin{equation}
\Gamma^{(1)} \equiv g_\varepsilon \hbar \int d^3\vec{x}\,{\rm div}\, j^{(0)}(\vec{x}) \varepsilon^{ (1)}(\vec{x}) \,.
\end{equation}

\ni In particular, in the case of Eqs. (37) and (38) with $c_{l\, m_l} = 0$ for $l>0$ the energy width (40) calculated for the system itself becomes the divergent expression

%rownanie 41
\begin{equation}
\Gamma^{(1)} = \frac{g^2_\varepsilon \lambda^2 \hbar}{4\pi c \tau_S^2}\, \lim_{\vec{x} \rightarrow \vec{x}_S} \frac{1}{|\vec{x}-\vec{x}_S|} \,,
\end{equation}

\ni containing the divergence typical for the classical selfenergy of a pointlike source. Of course, the pointlike character of a classical source is only an idealization, apparently not allowed in the discussion of selfenergy width $\Gamma^{(1)}$.

Thus, in stationary situations, the state vector (27) satisfying Eq. (24) takes the form

%rownanie 42
\begin{equation}
\psi^{(1)}(t) = \psi^{(0)}(t) \exp\left[ -\frac{1}{\hbar}\, \Gamma^{(1)}(t - t_0)\right] \;,
\end{equation}

\vspace{0.15cm}

\ni where $\psi^{(0)}(t_0) = \psi^{(1)}(t_0)$. Hence,

%rownanie 43
\begin{equation}
<\!\psi^{(1)}(t) | \psi^{(1)}(t)\!> = \exp \left[-\frac{2}{\hbar}\, \Gamma^{(1)}(t - t_0) \right]
\end{equation}

\vspace{0.15cm}

\ni for $<\!\psi^{(1)}(t_0)|\psi^{(1)}(t_0)\!> = 1$. 

\vspace{0.2cm}

\ni {\bf 6. The Sun changes norms of matter states on the Earth surface}

\vspace{0.2cm}

Notice that in chronodynamics, in consequence of absorption of Sun photons in the daily cycle from sunrise to sunset, the states of atoms and molecules of the matter on the Earth surface should have their norms changed. In fact, from Eqs. (16) and (9)

%rownanie 44 
\begin{eqnarray}
\Gamma(t) & = & - \frac{\hbar c}{\lambda^2} \int d^3\vec{x}\, \varepsilon(x)\left(\Box + \frac{1}{\lambda^2}\right) \varepsilon(x) \nonumber \\ & = &  \frac{\hbar c}{\lambda^2} \int d^3\vec{x}\, \left\{\left[\vec{\partial} \varepsilon(x)\right]^2 + \varepsilon(x) \left( \frac{1}{c^2}\, \frac{\partial^2}{\partial t^2} - \frac{1}{\lambda^2}\right) \varepsilon(x)\right\} \,,
\end{eqnarray} 

\vspace{0.15cm}

\ni where $\varepsilon(x)$ is excited by the source $\partial_\mu j^\mu(x)$ (during the day $\partial_\mu j^\mu(x) < 0$, as Sun photons are absorbed, implying mixed photon numbers in $\psi(t)$; then the radiative excitation of atoms and molecules transits mainly into the kinetic excitation of their neighbourhood, what leads to growing temperature). Since the field  $\varepsilon(x)$ is practically static (in some time intervals), Eq. (44) implies that during the day $\Gamma(t) > 0$ for $\lambda$ large enough, and then norms of matter states on the Earth surface should be decreased. In contrast,  for $\lambda$ small enough the effect should be opposite, but negligible, as then during the day $\Gamma(t) \propto g^2_\varepsilon \lambda^4 \simeq 0$, causing practically no change of norms of these matter states.

However, according to chronodynamics the Sun not only emits photons (and various matter particles such as electrons and neutrinos), but also excites the time deviations described by the solar field $\varepsilon^{\rm ex}(x)$ that for atoms and molecules on the Earth plays the role of an external field (practically static), leading to the energy width

%rownanie 45
\begin{eqnarray}
\Gamma^{\rm ex}(t) & \equiv & g_\varepsilon \hbar \int d^3\vec{x}\, \partial_\mu j^\mu(x) \varepsilon^{\rm ex}(x) \nonumber \\ & = & - \frac{\hbar c}{\lambda^2} \int d^3\vec{x}\, \varepsilon^{\rm ex}(x) \left(\Box + \frac{1}{\lambda^2}\right) \varepsilon(x) \,,
\end{eqnarray} 

\vspace{0.1cm}

\ni where $\partial_\mu j^\mu(x)$, detecting the solar field $\varepsilon^{\rm ex}(x)$ on the Earth, is at the same time the source exciting its own field $\varepsilon(x)$ [that appears also in Eq. (44)]. When solar photons are ab\-sorbed on the Earth surface, then $\partial_\mu j^\mu(x) <0$, while on the average $\varepsilon^{\rm ex}(x) >0$ or $\simeq 0$, respectively, for $\lambda$ large or small enough, as $\partial_\mu j^{\mu\,{\rm ex}}(x) > 0$ on the Sun. Thus, Eq. (45) shows that during the day $\Gamma^{\rm ex}(t) < 0$ or $\simeq 0$, respectively. The energy widths (44) and (45) are to be summed. Since the factor $\partial_\mu j^\mu(x)$ in the integrands of $\Gamma^{\rm ex}(t)$ and $\Gamma(t)$ is the same and the solar field $\varepsilon^{\rm ex}(x)$ should dominate over $\varepsilon(x)$ (even on the Earth), the energy width $\Gamma^{\rm ex}(t)$ should dominate over $\Gamma(t)$ leading, respectively, to an increase or practically no change of norms of matter states on the Earth surface during the day.

The change in time of norms of matter states may be translated into a change in time of the averaged matter density ({\it i.e.}, averaged number of matter particles contained in a unit volume; strictly speaking, for matter states with changing norms the particle number cannot be a "good"\, quantum number corresponding to a constant of motion, so {\it e.g.} a state of one hydrogen atom acted on by the Sun is here only an approximation, but an extremely good one). Of course, both changes are expected to be very small, since $\Gamma^{\rm ex}(t)$ and $\Gamma(t)$ are of the order $ g^2_\varepsilon$ where $ g_\varepsilon$ is tiny. The actual value of norms of matter states is always normalized to one, and only comparative observations corresponding, perhaps, to cosmological time scale could reveal its changes in time. Then, the lengthdimensional constant $\lambda $ of, perhaps, cosmological length scale might be here natural.

We would like to emphasize that, from the viewpoint of the overall quantum theory including hopefully the physical spacetime interacting on a quantum level, our conclusion on the change of norms of matter states is valid only for such matter states which are observed {\it alone}, without any accompanying states of quantum spacetime. Then, their interrelations with all quantum states of matter plus spacetime or of spacetime alone are described implicitly by the operator of their energy width multiplied by $-i$. In this argument we assume, of course, that the observation of matter states separated from quantum spacetime states is practically possible (like the observation of matter states, {\it e.g.} electrons, separated from energy states, {\it e.g.} photons).

In an extreme case, our optical-model-like theory might work even if its background theory, including the physical spacetime interacting on a quantum level, did not exist as a correct, selfconsistent theoretical scheme. In this case, an act of abstraction from the familiar optical-model mechanism  would be realized in application to the spacetime as a possible physical medium. Then, the change in time of the norms of matter states would get a fundamental character. Notice generically that, from the viewpoint of the time-temperature analogy, the energy width multiplied by $-i$, $-i\Gamma$, is an analogue of the heat $Q$ (what implies the complex internal energy $U$ in the first law of thermodynamics).

\vspace{0.2cm}

\ni {\bf 7. Detection of time deviations by a stationary source of particles}

\vspace{0.2cm}

Consider again a pointlike stationary system, located now at the point $\vec{x}_D$, producing a constant averaged current $\vec{j}^{(0)}(\vec{x})$ of particles of all sorts emitted at the rate $\tau_D$. Such a system excites according to Eq. (22) the static field $\varepsilon^{(1)}(\vec{x})$ of time deviations that, however, will not be discussed here. Let the system be situated in an external static 
field $\varepsilon^{\rm ex}(\vec{x})$ of time deviations, excited {\it e.g.} like the field $ \varepsilon^{ (1)}(\vec{x})$ described by Eq. (38) in Section 5. In such a case,

%rownanie 46
\begin{equation}
\varepsilon^{\rm ex}(\vec{x}) = \frac{g_\varepsilon \lambda^2}{4\pi c \tau_S}\, \frac{\cos (|\vec{x}-\vec{x}_S| /\lambda)}{|\vec{x}-\vec{x}_S|}\sum_{l\, m_l} c^{\rm ex}_{l\, m_l} Y_{l\,m_l}(\theta,\phi)\,,
\end{equation}

\ni where the mathematical model (37) is applied to $q^{\rm ex}(\vec{x}) \equiv {\rm div}\, \vec{j}^{\rm ex}(\vec{x})$.

In the present situation, the energy width (25) calculated for the system in external field is time-independent,
      
%rownanie 47
\begin{equation}
\Gamma^{(1)} \equiv g_\varepsilon \hbar \int d^3\vec{x}\,{\rm div}\, \vec{j}^{(0)}(\vec{x}) \varepsilon^{\rm ex}(\vec{x}) \,,
\end{equation}

\ni and the state vector (27) satisfying Eq. (24) can be presented as in Eq. (42). Hence, also Eq. (43) holds. If the mathematical model (37) is applied not only to $q^{\rm ex}(\vec{x}) \equiv {\rm div}\, \vec{j}^{\rm ex}(\vec{x})$ but also to $q^{(0)}(\vec{x}) \equiv {\rm div}\, \vec{j}^{(0)}(\vec{x})$, and if $c_{l\, m_l} = 0$ and $c^{\rm ex}_{l\, m_l} = 0$ for $l>0$, then

%rownanie 48
\begin{equation}
\Gamma^{(1)} = \frac{g^2_\varepsilon \lambda^2 \hbar}{4\pi c \tau_D \tau_S}\, \frac{\cos (|\vec{x}_D - \vec{x}_S| /\lambda)}{|\vec{x}_D - \vec{x}_S|} \,.
\end{equation}

\ni Note that

%rownanie 49
\begin{equation}
\Gamma^{(1)} \simeq \frac{g^2_\varepsilon \lambda^2 \hbar}{4\pi c \tau_D \tau_S}\, \frac{1}{|\vec{x}_D - \vec{x}_S|}
\end{equation}

\ni if $\lambda \gg |\vec{x}_D - \vec{x}_S|$, and $\Gamma^{(1)} \simeq 0$ if $\lambda \ll |\vec{x}_D - \vec{x}_S|$.

We can see that the appearance of energy width $\Gamma^{(1)} \neq 0$ in the state vector $\psi^{(1)}(t)$ describing the system with $q^{(0)}(\vec{x}) \equiv {\rm div}\, \vec{j}^{(0)}(\vec{x}) \neq 0$ is a signal coming from the external field $\varepsilon^{\rm ex}(\vec{x})$ of time deviations. Thus, in principle, such a system may play the role of a detector for the external time deviations $\delta t^{\rm ex}(\vec{x}) \equiv \varepsilon^{\rm ex}(\vec{x}) t$.

\vspace{0.2cm}

\ni {\bf 8. Excitation of time deviations by a vibrating source of particles}

\vspace{0.2cm}

Consider a pointlike vibrating system, located at the point $\vec{x}_S$, producing a changing averaged current $\vec{j}^{(0)}(x)$ of particles of all sorts emitted at the averaged rate $ 1/\tau_S $. It may be {\it e.g.} a source of radio waves. According to Eq. (22), this system is also a pointlike source of the varying field $\varepsilon^{(1)}(x)$ describing the time deviations $\delta t^{(1)}(x) \equiv \varepsilon^{(1)}(x)$ excited in such a way. In a simple model, the source $q^{(0)}( x) \equiv \,\partial_\mu\, j^{\mu\,(0)}(x)$ of $ \varepsilon^{(1)}(x)$ may be 

%rownanie 50
\begin{equation}
\partial_\mu \,j^{\mu\,(0)}(x) \equiv \frac{1}{\tau_S}\delta^3(\vec{x}-\vec{x}_S) \cos \omega_S t\,,
\end{equation}

\ni leading to Eq. (22) in the form

%rownanie 51
\begin{equation}
\left( \Box + \frac{1}{\lambda^2}\right) \varepsilon^{(1)} (x) = - \frac{g_\varepsilon \lambda^2}{c \tau_S} \,\delta^3(\vec{x}-\vec{x}_S) \cos \omega_S t\,.
\end{equation}

\ni Then,

%rownanie 52
\begin{equation}
\varepsilon^{(1)} (x) =  \frac{g_\varepsilon \lambda^2}{4\pi c \tau_S}\, \frac{\cos(\sqrt{\omega^2_S/c^2 + 1/\lambda^2} \,|\vec{x} - \vec{x}_S| - \omega_S t)}{ |\vec{x} - \vec{x}_S|} \,.
\end{equation}

In the present situation, the energy width (25) calculated for the system itself is the time-dependent divergent expression

%rownanie 53
\begin{equation}
\Gamma^{(1)}(t) = \frac{g^2_\varepsilon \lambda^2 \hbar}{4\pi c \tau_S^2}\, \lim_{\vec{x} \rightarrow \vec{x}_S} \frac{\cos \omega_S t}{|\vec{x}-\vec{x}_S|} \,,
\end{equation}

\ni caused by the pointlike character of the classical source. Obviously, this is only an idealization not working in the discussion of selfenergy width $\Gamma^{(1)}(t)$. Formally, the related divergent exponent appearing in the state vector (27) is

%rownanie 54
\begin{equation}
-\frac{1}{\hbar} \int^t_{t_0} dt'\, \Gamma^{(1)}(t') = -\frac{g^2_\varepsilon \lambda^2}{4\pi c \tau_S^2}\, \lim_{\vec{x} \rightarrow \vec{x}_S}\, \frac{\sin \omega_S t - \sin \omega_S t_0}{|\vec{x}-\vec{x}_S|} \,.
\end{equation}

In the spherical wave solution (52) for the field $\varepsilon^{(1)}(x)$ satisfying the tachyonic-type Klein-Gordon equation the wave number is $|\vec{k}| = \sqrt{\omega^2_S/c^2+1/\lambda^2} > \! \omega_S/c$, while in the case of conventional Klein-Gordon equation it would be $|\vec{k}| = \sqrt{\omega^2_S/c^2 - 1/\lambda^2} \equiv \sqrt{\omega^2_S/c^2 - (mc/\hbar)^2} \leq \omega_S/c$ for $\omega_S/c \geq 1/\lambda \equiv mc/\hbar$. Thus, the group velocity of the wave (52) is equal to $|\vec{k}|/(\omega_S/c^2) = c\sqrt{\omega^2_S + c^2/\lambda^2}\,/\omega_S > c$, so it is ultraluminal, whilst the group velocity in the case of conventional Klein-Gordon equation would be $|\vec{k}|/(\omega_S/c^2) = c\sqrt{\omega^2_S - c^2/\lambda^2}\,/\omega_S < c$ for $\omega_S \geq c/\lambda \equiv mc^2/\hbar$. Therefore, the field $\varepsilon^{(1)}(x)$ satisfying the tachyonic-type Klein-Gordon equation cannot transport the energy of a quantum system and so, any physical information (at least, in the sense ascribed conventionally to this term).

\vspace{0.2cm}

\ni {\bf 9. Final remarks}

\vspace{0.2cm}

The absence of energy transport by the field $\varepsilon(x)$ is generally evident in the framework of chronodynamics, where no observable energy related to $\varepsilon(x)$ is ascribed to the quantum system. Instead, the quantum system gets the energy width defined by the (trivial) operator ${\bf 1}\,\Gamma(t)$ with $\Gamma(t)$ related to the classical field $\varepsilon(x)$ through Eq. (16). In fact, any observable energy related to the field $\varepsilon(x)$ (and its interaction with particles) is absent from the state equation (13), even in the form of some (trivial) operator proportional to {\bf 1}, because all states of the physical spacetime, if treated dynamically on a quantum level, are ideologically projected out from the Hilbert space of the considered quantum system. It leaves in the state equation (13) the operator of energy width as their only remainder (according to the general projecting-out procedure in the Hilbert space [4]). Of course, the energy related {\it via} Eq. (9) to the classical (or, more precisely, classical-valued) field $\varepsilon(x)$,

%rownanie 55
\begin{equation}
 E_\varepsilon(t) =  \int d^3\vec{x}\,\left\{\frac{\hbar c}{2\lambda^2}\left[\frac{1}{c^2} \dot{\varepsilon}^2(x) + \left(\vec{\partial}\, \varepsilon(x)\right)^2 - \frac{1}{\lambda^2}{\varepsilon}^2(x) \right] + g_{\varepsilon} \hbar\, \partial_\mu j^\mu(x) \varepsilon(x) \right\}\,,
\end{equation}

\ni and the corresponding (trivial) Hermitian operator ${\bf 1} E_\varepsilon(t)$ can be introduced formally into the state equation (13) through the unitary transformation

%rownanie 56
\begin{equation}
\psi'(t) =\exp\left[ -\frac{i}{\hbar} \int^t_{t_0} dt' {\bf 1} E_\varepsilon(t') \right] \psi(t) = \psi(t) \exp\left[ -\frac{i}{\hbar} \int^t_{t_0} dt' E_\varepsilon(t') \right] 
\end{equation}

\ni which, in fact, is reduced to a phase transformation and thus, is physically unobservable. Naturally, the situation with the antiHermitian operator $-i{\bf 1} \Gamma(t)$ (also trivial) is quite different as it generates changes of the norm of state vector $\psi(t)$ which, in principle, are observable. This is so, in spite of the fact that $\Gamma(t)$ as given in Eq. (16) is formally equal to the interaction part of $E_\varepsilon(t)$ in Eq. (55) [of course, all the difference is caused by the factor $-i$ at $\Gamma(t)$].

However, in chronodynamics an ultraluminal communication {\it via} the classical field $\varepsilon(x)$ is allowed, resulting in an apparent transfer of energy width $\Gamma(t)$ between two quantum systems (for which the averaged four-currents of particles, {\it e.g.} photons, are not to be conserved locally {\it i.e.}, $\partial_\mu j^\mu(x) \neq 0$ for both). Such a transfer changes slightly the norms of state vectors of both systems what, in principle, can be observed in time.

\vfill\eject

~~~~
\vspace{0.5cm}

{\centerline{\bf References}}

\vspace{0.5cm}

{\everypar={\hangindent=0.6truecm}
\parindent=0pt\frenchspacing

{\everypar={\hangindent=0.6truecm}
\parindent=0pt\frenchspacing

[1]~{\it Cf. e.g.} A.L. Fetter and J.D.~Walecka, {\it Quantum Theory of Many-Particle Systems}, McGraw-Hill 1971.

\vspace{0.2cm}

[2]~W. Kr\'{o}likowski,  {\it Acta Phys. Pol.} {\bf B 24}, 1903 (1993); {\bf B 26}, 1511 (1995); {\bf B 27}, 2159 (1996); {\bf B 29}, 2081 (1998); {\it Nuovo Cim.} {\bf  107 A}, 1759 (1994); in the present paper, the description of time deviation and the structure of energy width are improved (as we hope).

\vspace{0.2cm}

[3]~P.A.M. Dirac, {\it The Principles of Quantum Mechanics}, 4th edition, Oxford University Press 1959.

\vspace{0.2cm}

[4]~For a general procedure of projecting out all states of an interacting subsystem from the Hilbert space of a larger quantum system {\it cf.} W.~Kr\'{o}likowski and J.~Rzewuski,  {\it Nuovo Cim.} {\bf  25 B}, 739 (1975).

\vspace{0.2cm}

[5]~S. Tomonaga, {\it Progr. Theor. Phys.} {\bf 1}, 27 (1946).

\vspace{0.2cm}

[6]~J. Schwinger, {\it Phys. Rev.} {\bf  74}, 1439 (1948).

\vspace{0.2cm}

\vfill\eject

\end{document}